
\input phyzzx
\overfullrule=0pt
\def\chat{\hat c}
\def\cht{{{\hat c \over 2}}}
\def\tto{\Theta_1(u-{1\over 2}|\tau) \Theta_1(u-{\tau \over 2}|\tau)
         \Theta_1(u+{{\tau+1}\over 2}|\tau)}
\def\ttt{\Theta_2(u) \Theta_3(u) \Theta_4(u)}
\def\half{{1\over 2}}

\def\pa{\partial}
\def\zt{\tilde Z}
\def\cmp#1{{\it Comm. Math. Phys.} {\bf #1}}
\def\pl#1{{\it Phys. Lett.} {\bf #1B}}

\def\np#1{{\it Nucl. Phys.} {\bf B#1}}

\def\jmath#1{{\it J. Math. Phys.} {\bf #1}}

\def\ct{{c\over 24}}
\REF\LVW{W. Lerche, C. Vafa and N.P. Warner, \np {324} 427 (1989) }
\REF\WIT{E. Witten, ``On the Landau-Ginzburg description of $N=2$ minimal
models", IAS preprint IASSNS-HEP-93/10}
\REF\ELG{
E. Witten, \cmp {109} 525 (1987);
K. Pilch, A. Schellekens and N. Warner, \np {287} 362 (1986);
O. Alvarez, T. Killingback, M. Mangano and P. Windey, \cmp {111} 1
(1987);
E. Witten, in {\it Elliptic curves and modular forms in algebraic
topology}, ed. P. Landweber (Springer Verlag 1988) }
\REF\ANM{A. Schellekens and N. Warner, \pl {177} 317 (1986);
A. Schellekens and N. Warner, \np {287} 317 (1986)}
\REF\DFY{P. Di Francesco and S. Yankielowicz, \np {409} 186 (1993) }
\REF\DFYO{P. Di Francesco, O. Aharony and S. Yankielowicz, ``Elliptic
genera and the Landau-Ginzburg approach to $N=2$ orbifolds", to appear
in {\it Nucl. Phys.} {\bf B}}
\REF\KYY{T. Kawai, Y. Yamada and S.-K. Yang, ``Elliptic genera and $N=2$
superconformal field theory", KEK preprint KEK-TH-362}
\REF\HEN{M. Henningson, ``$N=2$ gauged WZW models and the elliptic
genus", IAS preprint IASSNS-HEP-93/39}
\REF\Mer{A. Schellekens, \cmp{153} 159 (1993)}
\REF\GSW{M.B. Green, J.H. Schwarz and E. Witten, {\it Superstring
Theory} (Cambridge monographs on mathematical physics 1987)}
\REF\DGGH{D. Gepner, \cmp {142} 433 (1991) }
\REF\SchF{A. Schellekens, \np{366} 27 (1991)}
\REF\ScY{A. Schellekens and S. Yankielowicz, \np{334} 67 (1990)}
\REF\Eguc{T. Eguchi, H. Ooguri, A. Taormina and S.-K. Yang
\np{315} 193  (1989)}
\rightline{TAUP-2123-93}
\rightline{NIKHEF-H/93-27}
\rightline{hep-th/9311128}
\rightline{November 1993}
\vskip 1cm
\title{\bf Charge sum rules in N=2 theories}
\author {O. Aharony, S. Yankielowicz
\footnote{\dagger}{Work supported in part by
the US-Israel Binational Science Foundation and the Israel Academy
of Sciences.}}
\address{ School of Physics and Astronomy\break
Beverly and Raymond Sackler \break
Faculty of Exact Sciences\break
Tel Aviv University\break
Ramat Aviv, Tel-Aviv, 69978, Israel}
\author {A. N. Schellekens}
\address{NIKHEF-H \break
P.O. Box 41882 \break
1009 DB Amsterdam, The Netherlands}

\abstract{
We derive sum rules involving moments of the $U(1)$ charge in the Ramond
sector of $N=2$ super--conformal field theories. These charge sum rules
are obtained by analyzing the modular properties of the elliptic genus.
The lowest order sum rule, $<Q^2> = {\chat \over 12}$, pertains to the
average of the charge squared over the Ramond ground ring. The higher
sum rules contain information on the null state structure of the
underlying chiral algebra.
}
\endpage

\chapter{ Introduction}

Superconformal field theories (SCFTs) with $N=2$
have been the subject of a
large number of studies in the last few years, mainly due to their role
in the construction of space--time supersymmetric vacua in string theory.
Much is known about the structure of such theories, in particular for
unitary models, to which we will restrict ourselves here. Most of
the relevant information resides in an algebraic structure known as
the chiral ring, introduced in [\LVW], to which paper
we refer for details
and definitions.
Although the algebraic structure of $N=2$ SCFTs and in particular of
their chiral rings is tightly constrained, a complete classification
of all such  SCFTs is not yet available.

One of the important quantities
which characterize an $N=2$ SCFT is a torus partition function
defined as [\WIT]
$$ Z(q,z,{\bar z}) =
\Tr[(-1)^F q^{L_0-\ct} {\bar q}^{{\bar L}_0-\ct}
e^{2\pi i(zJ_0 + {\bar z} {\bar J}_0)}]\ , \eqn\defz $$
where
$L_0$ and $J_0$ are the Virasoro and $U(1)$ zero mode generators and
bars refer to rightmovers.
The trace is
over all states in the Ramond sector.
The computation
of the partition function for $N=2$ SCFTs is generally very difficult,
and possible only for simple cases like the $N=2$ minimal models.

The situation is better for the elliptic genus [\ELG,\ANM,\WIT],
which is the partition
function taken at ${\bar z}=0$,
$$ Z(z | \tau) = \Tr[(-1)^F q^{L_0-\ct} {\bar q}^{{\bar L}_0-\ct}
                   e^{2\pi izJ_0}].         \eqn\defzz $$
In this case, a simple supersymmetry
argument shows that all states of non--zero ${\bar L}_0-\ct$
cancel in pairs,
so that the elliptic genus is a holomorphic function of
 $q = e^{2\pi i \tau}$.
The elliptic genus is invariant under smooth deformations of the theory
which preserve the right--moving supersymmetry, and in particular it is
a topological invariant of the target space for supersymmetric sigma
models. At $z=0$ the elliptic genus equals the Witten index
$\Tr[(-1)^F]$, which gives the Euler characteristic in the case of sigma
models. The elliptic genus can quite easily be computed for
Landau-Ginzburg models [\WIT,\DFY] and for their orbifolds [\DFYO,\KYY],
and has also been calculated for some sigma models [\KYY] and for the
supersymmetric
$SU(2)/U(1)$ coset model [\HEN].

Another essential property of the elliptic genus is its behavior
under modular transformations.
This behavior can be derived using the
same argument
that was used in [\ANM] for the anomaly generating function.
First one observes
that
$Z(0,\tau)$ is obtained from a path integral on the torus
with periodic boundary conditions along both cycles, often referred
to as the ``PP-sector" (note that the operator $(-1)^F$ acts on left-
as well as on right-movers). Under $\tau \to -{1\over\tau}$
the characters in this sector
transform among themselves, and we consider a combination
of these characters so that $Z(0 | \tau)$ is modular invariant, for
example the diagonal invariant.
Since $Z(0 | \tau)$ is a constant this may seem trivial, but
modular invariance is in general
a non-trivial property when we write $Z$ in terms
of $N=2$ characters, and becomes even more interesting
when we include the $z$-dependence.
To do so, note that including
a   $z$-dependence for the
$U(1)$ characters affects their transformation
under $\tau \to -{1\over\tau}$
in only two ways: a rescaling of the argument
 $z$ by ${1\over\tau}$, and a factor
$e^{\pi i N z^2 / \tau}$,
where $N$ is the normalization of the $U(1)$ current,
$ J(z) J(w) ={N \over (z-w)^2}$ (the standard normalization in
$N=2$ models is $N=\chat \equiv {c\over 3}$).
Both effects are
the same for all characters, and hence they only affect the
partition function in a global way.
Hence we obtain (see also [\KYY]):
$$ \eqalign
{Z(z|\tau +1) &= Z(z|\tau) \cr
 Z({z\over \tau}|-{1\over \tau}) &= e^{\pi i\chat  {{z^2}\over{\tau}}}
 Z(z|\tau) \ . \cr} \eqn\modtrant $$
Holomorphic modular functions with similar transformations have
appeared at least twice before in the literature. The first was
the anomaly generating function of [\ANM], whose transformation
rule differs from \modtrant\ by a space-time dimension dependent
weight factor, and the second was the ``character valued" partition
of the meromorphic conformal field theories at $c = 24$ [\Mer]
whose partition
function is the absolute modular invariant $J(\tau)$. In both cases
the combination of holomorphicity and modular invariance allowed the
derivation of general rules governing the content of the theory. These
rules are respectively the Green-Schwarz factorization of space-time
chiral anomalies, and a relation between the levels and dual Coxeter
numbers of the Kac-Moody algebras forming a meromorphic $c=24$
modular invariant.

The main purpose of this paper is to investigate whether any such
general rules can be derived in the present case as well. We will
indeed find a universal
quadratic sum rule for the $U(1)$ charges of any $N=2$
theory. In addition we find many other sum rules, but they always
involve states of higher excitation levels, whose multiplicities
depend on the chiral algebra. Only for minimal models, whose
chiral algebra
is nothing but the $N=2$ algebra, can these sum rules be checked
directly, but of course nothing new can be learned there.
In all other cases the quadratic sum rule imposes a non-trivial
constraint on the chiral ring, and all other sum rules impose
rather complicated constraints on the action of the chiral algebra
on the Ramond ground states.

Another motivation for this work is a recent speculation that the
elliptic genus might be determined completely by the Poincar\'e polynomial.
Under an assumption regarding the analytic behavior
of the zeroes of the elliptic genus
at $\tau=0$ this was
shown  in [\DFY]. However, from the present analysis it follows that
uniqueness cannot be derived from modular invariance alone, indicating
that the assumption made in [\DFY] is not in general satisfied.
Nevertheless, we will find that deviations from the conjecture must be
such that they only affect sums of the twelfth power (or higher) of the
charges, and that all lower order sum rules are indeed determined by
the Poincar\'e polynomial.

In the next section we present the general parametrization of
the $N=2$ elliptic genus as dictated by modular invariance. Then,
in section 3, we derive the sum rules. In section 4 a different
analysis of the elliptic genus is presented that focuses on its zeroes,
and from which the quadratic sum rule is obtained in a different way.
In at least one case this method seems to be more powerful than
the one given in section 2.
In section 5 we discuss the application of the sum rule to
the special cases of Calabi-Yau, Landau-Ginzburg and
coset models. In the latter case we investigate
 the effects of
field identification fixed points.
Finally we discuss the restrictions
imposed by the sum rule on the null-vector structure of super W-algebras.

\chapter{Eisenstein expansion of the elliptic genus}

To remove the exponential factor in the transformation rule
\modtrant\ we consider the following modification
$$ \zt(z|\tau) = e^{\half \chat  z^2 G_2(\tau)} Z(z|\tau)  \eqn\ztdef
 $$
where $G_2(\tau)$ is the Eisenstein function defined by
$$ G_2(\tau) = 2\zeta(2)+2(2\pi i)^2 \sum_{n=1}^{\infty} \sigma(n) q^n
  \eqn\gtwodef $$
and $\sigma(n)=\sum_{d|n} d$. Then, since $G_2(\tau+1)=G_2(\tau)$ and
$G_2(-{1\over \tau})=\tau^2 G_2(\tau) -2\pi i\tau$, we find
that $\zt(z|\tau)=\zt(z|\tau+1)=\zt({z\over \tau}|-{1\over \tau})$.
Expanding $\zt$ now in a power series in $z$,
$$\zt(z|\tau)=
\sum_{k=0}^{\infty} z^{2k} D_{2k}(\tau), \eqn\ztexp $$
                                          and inserting the
transformations of $\zt$, we find that $D_{2k}(\tau)$ must be a modular
form of weight $2k$, meaning that $D_{2k}(\tau+1)=D_{2k}(\tau)$ and
$D_{2k}(-{1\over \tau})=\tau^{2k}D_{2k}(\tau)$. These transformation
properties, in  addition to the knowledge that these functions do not
have poles at $q=0$, allow us to determine them up to a few unknown
parameters. They can be expressed as polynomials in
the Eisenstein functions $G_4$ and $G_6$, for example
\input tables
\thicksize = 0pt
\thinsize = 0pt
\vskip 1.truecm
\begintable
\hfill $ D_0(q)$ &= $c_0$\hfill & ~~~~~~~~~~~ &
\hfill $ D_8(q)$~ &= $c_8 (G_4(q))^2$\hfill \nr
\hfill $ D_2(q)$ &= $0$\hfill &              &
\hfill $ D_{10}(q)$ &= $c_{10} G_4(q) G_6(q)$\hfill \nr
\hfill $ D_4(q)$ &= $c_4 G_4(q)$\hfill &     &
\hfill $ D_{12}(q)$ &= $c_{12}^1 (G_4(q))^3+c_{12}^2(G_6(q))^2$\hfill \nr
\hfill $ D_6(q)$ &= $c_6 G_6(q)$ \hfill &      &
\hfill $ D_{14}(q)$ &= $c_{14} (G_4(q))^2 G_6(q)\ ,$\hfill\endtable
\noindent
etc. Functions with higher weights all have at least two parameters, and
the number of parameters continues to increase in an obvious way.

\chapter{Sum rules}
The foregoing results enable us to express $Z(z | \tau)$ in terms of
a set of real parameters $c_{2k}$. By expanding $Z$ in $z$ and/or
$q$, and considering linear combinations from which these
parameters cancel, we obtain sum rules for certain powers of the
charges. Obviously $c_0 = \Tr [ (-1)^F] $, the Witten index.
{}From the vanishing of
$D_2$ we get
$$ \Tr[(-1)^F q^{L_0-{c\over{24}}} {\bar q}^{{\bar L}_0-{c\over{24}}}
J_0^2] = {{\chat E_2(q)}\over {12}} \Tr [(-1)^F] \ ,
   \eqn\genres $$
where for convenience we define normalized Eisenstein functions
$E_{2k}$ whose constant term is equal to 1. Then $G_{2} =
{\pi^2\over 3} E_2$, and $E_2(q) = 1 - 24 q + \ldots$.
To leading order in $q$ we find thus the
following sum rule for the charges of the Ramond ground states :
$$ \Tr[(-1)^F J_0^2 ] = {\chat \over 12} \Tr [(-1)^F ] \ .\eqn\SumRule $$
This is the main result of this paper.
If we apply it to the diagonal invariant of an $N = 2$ conformal
field theory whose Ramond ground states have multiplicity
one\foot{This means each Ramond module has at most
one state with $h=\ct$. This is true for example in coset theories as
long as
one does not extend their chiral algebra by, for example, simple
currents. If there are more Ramond ground states per module, their
relative sign of $(-1)^F$ is relevant, and then the more general
formula \SumRule\ should be used.}
the operator $(-1)^F$ acts in the same way on left- and
right-movers, and we get the interesting relation
$$ < J_0^2 > = { \chat \over 12}\ , $$
where the average is over the chiral ring.
This simplified form of the sum rule is valid, for example,
for any coset model, and in particular for the
minimal models where it can be
verified easily.
This sum rule is quite surprising since in unitary $N=2$
theories the lowest and highest charges are plus or minus $\chat \over 2$
so that one
would perhaps expect $<J_0^2>$ to be quadratic in $\chat$, but it
turns out to be linear.

Note that \genres\ upon expansion in $q$ yields a quadratic sum rule
for every excitation level. It is perhaps instructive to compare this
result with the analogous situation for chiral anomalies. In that case
the argument is very similar, but first of all the presence of an extra
weight factor in the modular transformations shifts the order of $J_0$
upward, so that one gets relations among higher order traces.
Secondly,
the elliptic genus of a superstring can have a pole at $q=0$. In that
case the analog of $D_2$ does not vanish, but it is still true that its
constant term vanishes\rlap.\foot{Furthermore
the full $q$-dependence can
be determined from the residue of the pole.}
This observation was sufficient to
prove Green-Schwarz factorization for the chiral anomalies in string
theory [\ANM].

\def\cN{{\cal N}}
Returning now to $N =2 $ models, let us consider
the expansion of \genres\ to higher orders in $q$. For the first
excited level one obviously gets
$$ \Tr [ (-1)^F J_0^2 ]_{\rm level 1} = -2 \chat \cN \ , $$
where $\cN = \Tr[(-1)^F]$
denotes the total number of Ramond ground states. In this
case the $(-1)^F$ factor clearly {\it is} important, since some of the
states at the first level are created by the $N=2$ supercurrent,
and hence contribute with the opposite sign.
It is
interesting to inspect this sum rule for the special case of the
minimal $N=2$ models. Minimality implies that excitations can only be
produced by acting with generators of the $N=2$ algebra itself.
Acting on a generic Ramond ground state with charge $Q$  and $(-1)^F=1$,
the supercurrents
produce two states with $(-1)^F = -1$           and charges $Q\pm1$,
except for the maximal and minimal charge states
(those
with $Q = {\chat\over 2}$). For the latter two states one can show,
using the $N=2$ algebra, that the states with charge
$\pm |{\chat \over 2}+1|$
are in fact null states. Because of supersymmetry
these $(-1)^F=-1$ states are matched by an equal number of
states with $(-1)^F=1$, generated from the ground states by $L_{-1}$
and $J_{-1}$ (modulo null states). These states have thus the same
charge as the ground state, $Q$. Hence the sum rule reads
$$ \sum_i [ 2 Q_i^2 - (Q_i - 1)^2 - (Q_i + 1)^2 ] -
2 [ (\cht)^2 -(\cht + 1)^2 ] = - 2 \chat \cN\ . $$
Here the sum is over the entire chiral ring, and the last term removes
the null states. This sum rule leads to the following relation between
$\chat$ and $\cN$:
$$ \chat = { \cN - 1 \over \cN + 1 }\ . $$
This relation is indeed satisfied for minimal models (note that
$\cN = k+1$). Conversely, this is a very simple and
direct way of obtaining
the set of allowed central charges for minimal unitary $N=2$ models.
When we apply the sum rule to non-minimal models
it yields a constraint on the squares of the charges of those
excited states that are created by the operators that extend the
chiral algebra.

One can also consider higher powers of $J_0$. After some straightforward
algebra, we obtain the following sum rule for $J_0^4$:
$$ \Tr [ (-1)^F q^{L_0-\ct} \bar q^{\bar L_0 -\ct}
J_0^4 ] = c_4 E_4(q) + {1\over 48} \chat^2 (E_2(q))^2 \Tr[(-1)^F]
\eqn\Quartic $$
(here $c_4$ has been redefined by an irrelevant factor). The unknown
constant $c_4$ can be determined at level 0, and once it is known
we can predict the quartic traces at all the higher levels. In a
similar way slightly more complicated sum rules for traces of order
6, 8 and 10 can be derived, as well as for a particular linear
combination of traces of order 14 and 12.

\chapter{Analysis of the zeroes of the elliptic genus}
In this section we present a different way of analyzing the
elliptic genus, by parametrizing the zeroes in the first argument, $z$.
In the theories we are interested in, the $U(1)$ charges
of all states are (in the Neveu-Schwarz sector)
                  integer multiples of ${1\over h}$ for some integer $h$,
and we will find it convenient to change variables to $u = z/h$, and to
define $t=e^{2\pi iu}$. Obviously, $\chat h$ is then an integer, since
the highest charge state has charge $\chat$ in this sector.

The modular transformations of the elliptic
genus (which we will now consider as a function of $u$ and $\tau$) are
as follows.
    The transformations in the $\tau$ variable follow directly from
\modtrant:
$$ \eqalign
{Z(u|\tau +1) &= Z(u|\tau) \cr
 Z({u\over \tau}|-{1\over \tau}) &= e^{\pi i\chat h^2 {{u^2}\over{\tau}}}
 Z(u|\tau).\cr}\   \eqn\modtrantTWO $$
Using the last equality twice we can easily get
the charge conjugation symmetry
$$ Z(-u|\tau) = Z(u|\tau). \eqn\dual $$

The $u$ transformations of the elliptic genus are [\KYY]:
$$ \eqalign { Z(u+1|\tau) &= (-1)^{  \chat h} Z(u|\tau) \cr
 Z(u+\tau|\tau) &= (-1)^{  \chat h} e^{-\pi i\chat h^2 (\tau+2u)}
 Z(u|\tau).}  \eqn\modu $$
The Poincar\'e polynomial of the theory is defined as
$P(t) = \lim_{q\to 0} Z(u|\tau) $ and receives
contributions only from the ground ring of the $N=2$ theory.

To analyze the general form of the elliptic genus, we will use the Jacobi
theta functions, defined as in [\GSW].
Our first claim is that the elliptic genus of an $N=2$ theory may
always be written as
$$ \eqalign { Z(u|\tau) = &A(\tau) (\tto)^p \times \cr
&\prod_{i=1}^n \Theta_1(u-a_i(\tau)|\tau)
\Theta_1(u+a_i(\tau)|\tau)} \eqn\forma $$
where $p=1$ if $\chat h^2$ is an odd integer and $p=0$ otherwise,
$n={1\over 2}(\chat h^2 -3p)$ and $A(\tau)$ and $a_i(\tau)$ are some
functions (obviously the $a_i(\tau)$ are zeroes of the elliptic
genus).
We will assume that the zeroes are
well-behaved functions of $\tau$ (except perhaps in singular limits
such as $\tau \to 0$), a reasonable assumption since the function
$Z(u|\tau)$ is well-behaved.
Because of equation \dual  , if $a$ is a zero of $Z$ for some
$\tau$ then so is $-a$ (defined modulo $1$ and $\tau$). Therefore, all
zeroes of the elliptic genus come in pairs except perhaps for zeroes at
the half-integer lattice points
$0$, ${1\over 2}$, ${\tau \over 2}$ and ${{\tau+1}\over 2}$.
Double zeroes at these points can also be added to the $a_i$'s (a zero
at $u=0$ must be of even order because of the charge conjugation
symmetry).
However, if there is a single zero at one of these points for some
$\tau$, there must be a single zero there in a neighborhood of this
$\tau$ since it cannot move and still obey the $u\to -u$ symmetry.
Therefore, assuming the zeroes are well behaved we get that there must
be such a zero for all $\tau$.
We know that
for all $\tau$, $Z(0|\tau) = \Tr[(-1)^F]$ is the Witten index.
A zero at one of the other half-integer lattice points for all $\tau$
necessarily gives a zero at all three points, since $Z({\tau \over 2}|
\tau) = e^{-\pi i\chat h^2 {\tau \over 4}} Z({1\over 2}|-{1\over \tau})$
and $Z({{\tau+1}\over 2}|\tau) = Z({{\tau+1}\over 2}|\tau+1)$.
Therefore, either for all $\tau$ there is a single zero (after removing
double zeroes) at all three half-integer lattice points, or there is no
single zero at any of them for all $\tau$.

Let us now show that the total number of zeroes is always equal
to $\chat h^2$. For every $\tau$, $Z(u|\tau)$ is an analytic function
of $u$, so that by Cauchy's theorem the number of zeroes of $Z$ inside
the parallelogram connecting the points $(0,1,\tau+1,\tau)$ is ${1
\over {2\pi i}}$ times the $du$ integral of
${{Z'(u|\tau)}\over {Z(u|\tau)}}$
along the sides of the parallelogram ($Z'$ denotes the derivative of $Z$
with respect to $u$). However, from equation \modu\ it is easy to see
that ${{Z'(u+1|\tau)}\over{Z(u+1|\tau)}} = {{Z'(u|\tau)}\over{Z(u|\tau)}}
$ and ${{Z'(u+\tau|\tau)}\over{Z(u+\tau|\tau)}} =
{{Z'(u|\tau)}\over{Z'(u|\tau)}} - 2\pi i\chat h^2$ and therefore the
integral can easily be computed (since the parallel sides cancel up to
a constant) and equals $2\pi i\chat h^2$, and therefore $Z(u|\tau)$
indeed has $\chat h^2$ zeroes inside the parallelogram
                             for every $\tau$.
The claim \forma\ is now obvious if we define the $a_i(\tau)$ to be
the zeroes which are not at half-integer lattice points (choosing one
from each pair of such zeroes).
Since both sides of equation \forma\
(taking the right side without $A(\tau)$) have
the same zeroes and
no poles and furthermore
the same $u\to u+1$ and $u\to u+\tau$ transformations, their
ratio is (for every $\tau$) an elliptic function with no zeroes or
poles and therefore a constant, which we denote by $A(\tau)$.
Actually, in proving that the $u$ transformations of both
sides are the same we need to use the equality $(-1)^{\chat h} =
(-1)^{\chat h^2}$, meaning that if $\chat h$ is odd then $h$ must be
odd as well. This is correct, since if $\chat h$ is odd we get from
\dual\ and \modu\ that $Z({1\over 2}|\tau)=0$. Therefore $Z$ has
a zero at the half-integer lattice points, $p=1$ and $\chat h^2$ which
is the total number of zeroes must be odd as well  (the zero at $u=\half$
must be of odd order since $Z(\half+u|\tau)=Z(-\half-u|\tau)=
-Z(\half-u|\tau)$).    It is not clear
how this constraint may be obtained
directly from the $N=2$ algebra.

The next step of our proof uses well-known identities of theta functions
which state that
$$ \Theta_1(u-a_i(\tau)) \Theta_1(u+a_i(\tau)) =
   (\Theta_4)^{-2} [ \Theta_1(u)^2
                       \Theta_4(a_i(\tau))^2
   - \Theta_4(u)^2
   \Theta_1(a_i(\tau))^2]   \eqn\ident $$
(where we define $\Theta_i(u) = \Theta_i(u|\tau)$ and $\Theta_i =
\Theta_i(0|\tau)$)
to transform \forma\ into
$$ Z(u|\tau) = (\ttt)^p
        \sum_{k=0}^n A_k(\tau) (\Theta_1(u))^{2k}
       (\Theta_4(u))^{2(n-k)}
       \eqn\formb $$
where all functions depending only on $\tau$ and not on $u$ were
swallowed into the arbitrary functions $A_k(\tau)$.

Next we shall analyze the $\tau \to \tau+1$ and $\tau \to -{1\over \tau}$
transformations of \formb, and get the transformations of the $A_k(\tau)$
functions.
Let us first look at the $\tau \to \tau+1$ transformation, under which
$Z(u|\tau)=Z(u|\tau+1)$.
Inserting the
$\tau$ transformations
of the theta functions we get
$$ \eqalign { Z(u|\tau) = &Z(u|\tau+1) =\cr
 = &(\ttt e^{{1\over 4}i\pi})^p
   \sum_{k=0}^n A_k(\tau+1) (\Theta_1(u))^{2k} e^{\half i\pi k}
   (\Theta_3(u))^{2(n-k)}. \cr} \eqn\tranaa $$
Now, by using the identity
$$ (\Theta_3(u))^2 = (\Theta_4)^{-2} [(\Theta_4(u))^2 (\Theta_3)^2 -
  (\Theta_1(u))^2 (\Theta_2)^2 ]   \eqn\thetsq $$
and equating the coefficients on both sides, we find that the
transformation of $A_k$ is given by
$$ A_k(\tau) =\sum_{j=0}^k A_j(\tau+1) e^{{1\over 4}i\pi (p-2j)}
    ({{n-j}\atop {n-k}}) (-1)^{k}
    (\Theta_4)^{2(j-n)} (\Theta_2)^{2(k-j)} (\Theta_3)^{2(n-k)}.
    \eqn\tranad $$

The computation of the $\tau \to -{1\over \tau}$ transformation is
similar. After some algebra,      using the identity
$$ (\Theta_2(u))^2 = (\Theta_4)^{-2} [(\Theta_4(u))^2 (\Theta_2)^2
    - (\Theta_1(u))^2 (\Theta_3)^2 ],  \eqn\thetatwo $$
we get
the transformation equation
$$ A_k(\tau) =\sum_{j=0}^k A_j(-{1\over \tau})
    (-i\tau)^{\half \chat h^2} ({{n-j}\atop {n-k}}) (-1)^k
    (\Theta_4)^{2(j-n)} (\Theta_3)^{2(k-j)} (\Theta_2)^{2(n-k)}.
    \eqn\tranb $$
Note that in both transformations of $A_k$ only $A_j$'s with $j$ less than
or equal to $k$ participate.

Let us now start by looking at $A_0(\tau)$ : this is easily determined
by taking $u=0$, and we find that
$$ Z(0|\tau) = (\Theta_2 \Theta_3 \Theta_4)^p
           A_0(\tau) (\Theta_4)^{2n}\ . \eqn\zzero $$
Now, using the fact that
$Z(0|\tau) = \Tr[(-1)^F]$ for all $\tau$,        we find that
$$ A_0(\tau) = \Tr[(-1)^F] (\Theta_4)^{-2n}
    (\Theta_2 \Theta_3 \Theta_4)^{-p}.
    \eqn\azero $$
Hence $A_0$ is the same up to a multiplicative constant
in all theories with the same $\chat$, $h$, and
is obviously determined by the $q\to 0$ limit of the elliptic genus
(the Poincar\'e polynomial)
alone. It can be checked that the expression \azero\ obeys the equations
\tranad\ and \tranb\ for the case of $k=0$.
If $\Tr[(-1)^F]=0$ we find that $A_0(\tau)=0$.

In general the $q\to 0$ limit of equation \formb\ can be computed to be
$$ \lim_{q\to 0} Z(u|\tau) = (2 \cos(\pi u))^p
   \sum_{k=0}^n 2^{2k} (\sin(\pi u))^{2k}
   \lim_{q\to 0} (A_k(\tau) q^{{k\over 4}+{p\over 8}})
   \eqn\limz $$
and therefore, since we are dealing with theories with finite Poincar\'e
polynomials, the low $q$ behavior of the functions $A_k$ is of the form
$$ A_k(\tau) = C_k q^{-{p\over 8}-{k\over 4}} (1 + O(q))
\ . \eqn\lima $$
The Poincar\'e polynomial is then
$$ P(t) = (2\cos(\pi u))^p \sum_{k=0}^n C_k 2^{2k}
      (\sin(\pi u))^{2k}.  \eqn\poinc $$
Note that there are $\half \chat h^2$ terms in this expansion, but in
unitary theories there are only $\half \chat h$ non--zero terms in
the Poincar\'e polynomial, so that for $h$ not equal to one the higher
coefficients $C_k$ all vanish.

\section{Constraints on the chiral ring from the elliptic genus}

Let us now try to solve equations \tranad\ and \tranb\
for the $A_k$'s. The solution
we found above for $A_0$ is of course unique, both because of the
Witten index argument and also directly since the ratio $B$ of any other
solution divided by \azero\ would have to satisfy $B(\tau)=B(\tau+1)=
 B(-{1\over \tau})$ and have no poles, and therefore must be a constant.
The equations for the other $A_k$'s are linear, so that their general
solution is the sum of a special solution of the equations, plus a
general solution to the homogeneous equations obtained by setting all
$A_j$'s with $j<k$ to zero in equation \tranb. Looking     at these
homogeneous equations
$$ \eqalign {
A_k(\tau) &= e^{{1\over 4}i\pi (p+2k)} (\Theta_4)^{2(k-n)} (\Theta_3)^{2(n-k)}
    A_k(\tau+1) \cr
A_k(\tau) &=
    (-i\tau)^{\half \chat h^2}                      (-1)^k
    (\Theta_4)^{2(k-n)}                     (\Theta_2)^{2(n-k)}
     A_k(-{1\over \tau}) \cr }
  \eqn\homequ $$
we see that the equation for $A_k$ is very similar to the equation for
$A_0$, taken with $n-k$ instead of $n$. So, let us denote the unique
solution of that equation by
$$ F_n(\tau) = (\Theta_4)^{-2n} (\Theta_2 \Theta_3 \Theta_4)^{-p}
     \eqn\fn $$
and look at the ratio $B_k(\tau) =   A_k(\tau) /       F_{n-k}(\tau)  $.
The equations it satisfies are
$$ \eqalign { B_k(\tau) &= B_k(\tau+1) e^{\half i\pi k} \cr
              B_k(\tau) &= B_k(-{1\over \tau}) (i\tau)^k \cr } \eqn\bk
              $$
and the solution should behave as
                 $q^{-{k\over 4}}$ as  $q\to 0$ for consistency.
Let us now define $H_k(\tau) = (\eta(\tau))^{6k} B_k(\tau)$. The
transformations
of $H_k$ are now easily found to be
$$ \eqalign { H_k(\tau+1) &= H_k(\tau) \cr
              H_k(-{1\over \tau}) &= \tau^{2k} H_k(\tau) \cr } \eqn\hk $$
so that $H_k$ is a modular function of weight $2k$, and since it has no
poles at finite $\tau$ it must be a sum of Eisenstein functions as in
chapter 2. The $H_k$ functions then also have the correct $q\to 0$ behavior
because
of the $\eta(\tau)^{6k}$ factor. From the fact that there exist no non--zero
modular forms of weight $2$ we find that
$H_1$ must be zero, so that there is no
non--trivial solution to the homogeneous
equation for $A_1$, and $A_1$ is completely
determined by $A_0$ (their ratio is a constant function).

Let us examine the consequences of this fact -
it implies a linear restriction on the
Poincar\'e polynomial of such theories, since the ratio ${C_1}\over{C_0}$
(with the coefficients $C_i$ as defined above appearing in the expression
for the Poincar\'e polynomial) must also be a constant for every $\chat$
and $h$. In fact it is simpler to expand the Poincar\'e polynomial in a
power series in $u$, $P(t)=c_0+c_1(\pi u)^2+c_2(\pi u)^4+ \cdots$,
and then we found that $c_1/c_0$  must be the same for all theories with
the same transformations.
We can relate this ratio       to the average of $J_0^2$ in the chiral
ring. We defined
 $$ \Tr[(-1)^F e^{2\pi iuhJ_0}] = c_0 + c_1(\pi u)^2 + \cdots  \eqn\poin
$$ and
then by taking $u=0$ we immediately find $\Tr[(-1)^F] = c_0$, while
$$ \Tr[(-1)^F J_0^2] = -{1\over {4h^2}} {{\pa^2}\over {\pa(\pi u)^2}}
\Tr[(-1)^F e^{2\pi iuhJ_0}] = -{{c_1}
    \over {2h^2}}   \eqn\poinctwo              $$ (where the derivative
is taken at $u=0$). Now it is clear that the resulting relation
must be precisely \SumRule. This allows us to determine the ratio of
$c_1$ and $c_0$:
${{c_1}\over{c_0}}           = -{{\chat h^2}\over
6}$.

The analysis in this section has thus provided us with a second
method for deriving the quadratic sum rule (and in fact all other sum rules
as well).
It is clear that the two methods
are closely related, since they both give expansions in terms of Eisenstein
functions. Indeed, by expanding \formb\ in $u$  one can directly
relate the functions $A_k$ to the functions $D_{2k}$ introduced in
section 2. The foregoing discussion shows that the number of free
parameters is identical for each order $k \leq n$.
Nevertheless there is an important difference
between the two approaches: \formb\ expresses the elliptic genus in terms
of a finite number of functions $A_k$, whereas the sum in \ztexp\ is
infinite. This is a consequence of the fact that in the present section we
put in some extra information, namely the exact quantization of the $U(1)$
charge in units of ${1\over h}$.
In the approach of section 2 the same information would eventually
also emerge, because for a given charge quantization there is only a finite
number of independent traces. Hence for a sufficiently large power $k$,
$\Tr (-1)^F J_0^{k}$ is fully determined by the lower order traces, and hence
there are no more free parameters. However, it seems nearly
impossible to carry this out
in practice and demonstrate explicitly -- without using \formb\ --
that the number of free parameters in \ztexp\
is actually finite.
The number of parameters in the second expansion is exactly
the number of linearly independent modular forms of degrees less than or equal
to $2n=\chat h^2-3p$. For each degree one parameter may be determined from the
Poincar\'e polynomial.

\chapter{Applications}
\section{Calabi-Yau compactifications}
Looking at specific theories, the fact that $<J_0^2> = {{\chat}\over{12}}
$ has various implications. We first consider
Calabi-Yau models, which have $\chat=d$ (the complex dimension) and
charges quantized according to $h=1$ ($h$ is defined in chapter 4). The
total contribution to $\Tr[(-1)^F]$ from
states of charge $p-{d\over 2}$
is given by $N_p = \sum_{q=0}^d
(-1)^{p+q} h_{p,q}$ where $d=\chat$ and $h_{p,q}$ are the Hodge numbers
of the surface. The sum rule yields the strange relationship
$$ \sum_{p,q=0}^d h_{p,q} (-1)^{p+q} (p-{d\over 2})^2 =
    {d\over 12} \sum_{p,q=0}^d h_{p,q} (-1)^{p+q} = {d\over 12} \chi(M)
   \eqn\rescy $$
where $\chi$ is the Euler characteristic of the surface. This can be
used, with the help of Poincar\'e duality, \ie\ $N_p=N_{d-p}$, to
determine the Poincar\'e polynomials for $d=1,2$ and $3$ almost
completely.

For $\chat = 1$ the sum rule reads
${1\over 2}  N_0 =  {1\over 6} N_0$, so that $N_0=0$. This
reflects the well-known fact that the only 1-dimensional ``Calabi-Yau"
manifold is a torus.

For $\chat=2$ we find $2 N_0 = {1\over 6} (2 N_0 + N_1)$, which
immediately results in all Poincar\'e polynomials
being proportional to the one of the $K3$ surface,
$t^{-1} (2 + 20t + 2t^2)$. The overall factor is of course never
determined by considerations of modular invariance.
It is interesting to note that the same
result can be obtained from the requirement of Green-Schwarz
factorization of the chiral gauge and gravitational
anomalies in six-dimensional strings.
This factorization was derived in [\ANM] from a different,
but closely related elliptic genus, the character valued chiral
partition function. This function contains at least the same
information as the elliptic genus considered in the present
paper, since the $U(1)$ charge of the $N=2$ algebra is absorbed in
the $E_7$ gauge group, and contributes to the chiral anomaly.
The computations yielding the spectrum are a bit
more cumbersome, since they involve quartic traces rather than quadratic
ones, but on the other hand one can get a little more information from
the gravitational anomalies, which
determine the number of gauge singlets in the string spectrum.
In this approach,
the overall normalization of the Poincar\'e polynomial is fixed by
space-time supersymmetry considerations, \ie\ the number of gravitinos
in the spectrum.

For
$\chat=3$ we get
the relation ${9\over 2} N_0 + {1\over 2} N_1 = {1\over 2} N_0 +
{1\over 2} N_1$. This  implies that $N_0=0$,
so that
all Poincar\'e
polynomials must be proportional to $t^{-\half} (1 + t)$.
This is again not an unknown result, but the derivation is
quite interesting.

In all these cases (or generally
whenever $\chat h^2$ equals $1$\rlap,\foot{For $\chat h^2 = 1$ the
the analysis of chapter 4 shows that $p=1$ and hence $n=-1$, which
does not seem to make sense. However in this case the elliptic genus
vanishes identically, and therefore the analysis of section 4 does not
apply. The Eisenstein expansion and hence the sum rules remain valid,
however.}
$2$, $3$ or $5$) $n$ in equation
\formb\ is less than $2$, so that there is no freedom in determining
the elliptic genus.
For example,
in the $\chat=3$ Calabi-Yau case we find that the
elliptic genus is given by
$Z(z|\tau)=\half \Tr[(-1)^F] {{\Theta_1(2z|\tau)}\over
{\Theta_1(z|\tau)}}$ (this follows from the same
argument as given in [\DFY]; for a different approach see [\Eguc]).
For larger $\chat$, the elliptic genus is not
determined just from the modular
transformations.
Note that for $\chat=5$
the quadratic sum rule itself is not sufficient.
Furthermore for $\chat=2,3$ and 5 it is not at all manifest that
the higher order sum rules determine the elliptic genus.
In this case the
analysis of section 4, that uses charge integrality in the Neveu-Schwarz
sector, appears to be
more powerful than the sumrules alone. Nevertheless
careful
analysis of the higher sum rules should lead to the same conclusion.

\section{Landau-Ginzburg theories}
                    The
elliptic genus of Landau-Ginzburg theories, is known [\WIT,\DFY]
to be of the
form\break   $\prod_{i=1}^l {{\Theta_1((d-d_i)u|\tau)} \over
   {\Theta_1(d_iu|\tau)}}$,
where                                    $\chat h=\sum_{i=1}^l (d-2d_i)$,
and, as before, $u=z/h$ (in this case $h=d$).
The       $q\to 0$ limit is $\prod_{i=1}^l {{\sin((d-d_i)\pi u)} \over
{sin(d_i\pi u)}}$, and expanding  in a power series in $\pi u$ we find
that it behaves as
$$(\prod_{i=1}^l {{d-d_i}\over{d_i}}) (1 - {1\over 6}
(\pi u)^2 \sum_{i=1}^l ((d-d_i)^2 - d_i^2) + ...)    \eqn\lgone $$
which equals
$$(\prod_{i=1}^l {{d-d_i}\over{d_i}}) (1 - {1\over 6}
(\pi u)^2 \chat h^2 + ...)\ .\eqn\lgtwo $$
This trivially satisfies the quadratic sum rule. This is not surprising
since the expression for the elliptic genus is manifestly
modular invariant. Hence the sum rules do not yield any useful
information about such models.

\section{Coset models}
In coset models the sum rule  gives a non--trivial equality between
expressions involving the weights of the groups involved. In such models
the equation \SumRule\ may also be used to constrain the possible modular
invariants of a group $G$ - the diagonal elements appearing in such a
modular invariant must satisfy $<J_0^2> = {{\chat} \over {12}}$ for every
coset $G/H$ (where $J_0$ is the $U(1)$ charge of that coset for the
relevant representations and $\chat$ is its central charge).
For $G/H$ coset models it is natural    to take $h=k+g$ where $g$ is the
second Casimir invariant of $G$, and then the central charge is given by
[\DGGH]
$$ \chat h=h{{d_G-d_H}\over 2} -4\rho_G \cdot (\rho_G-\rho_H).\eqn\forc$$
The primary chiral states of these models are in a one to one
correspondence with pairs $(\Lambda,\omega)$, where $\Lambda$ is a weight
of the group $G$ at level $k$ and $\omega \in W(G/H)$ (meaning that
$\omega$ is a Weyl transformation of $G$ that preserves the positivity of
the positive roots of $H$), identified by the action of the outer
automorphism $\sigma$ of the algebra $G$  [\LVW,\DGGH].
When this automorphism has no fixed points, all of its orbits have a
length $|Z(G)|$ (where $Z(G)$ is the center of the group $G$), so that
ignoring the identification would just cause us to count each state
$|Z(G)|$ times and would not change $<J_0^2>$. The charge of the state
corresponding to $(\Lambda,\omega)$ in the Neveu-Schwarz sector is
given by [\DGGH]
 $$ hQ_{\omega}^{\Lambda} = hl(\omega)+2(\omega^{-1}(\Lambda+\rho_G)
   -\rho_G) \cdot (\rho_G-\rho_H)  \eqn\charge $$
where $l(\omega)$ is the length of the Weyl transformation $\omega$.
The charges in the Neveu-Schwarz sector are shifted by $\chat /2$
 relative to those in the Ramond sector, leading to $<Q^2>={\chat \over
{12}} + {{\chat^2}\over 4}$ in this sector. Inserting the formulas above,
and denoting by $N^G_k$ the number of weights of $G_k$, we find for the
case in which $\sigma$ has no fixed points the formula
$$ \eqalign{
   \sum_{\Lambda \in G_k} \sum_{\omega \in W(G/H)} (hl(\omega) +
 &2(\omega^{-1}(\Lambda+\rho_G)-\rho_G)   \cdot (\rho_G-\rho_H))^2 = \cr
  =N^G_k {{|W(G)|}\over{|W(H)|}}[&{h\over 12} (h{{d_G-d_H}\over 2} -
    4\rho_G \cdot (\rho_G-\rho_H)) + \cr
   &{1\over 4}(h{{d_G-d_H}\over 2}-4\rho_G \cdot (\rho_G-\rho_H))^2].\cr}
   \eqn\resgh $$
This formula may easily be checked to be correct in the $k=1$ case, where
the Poincar\'e polynomial is given by [\LVW,\DGGH]
$$ t^{\chat \over 2} P_{G/H}(t) = \prod_{\alpha \in \Delta_{G/H}}
   {{1-t^{k+g-\alpha \cdot \rho_G}}\over{1-t^{\alpha \cdot \rho_G}}}.
  \eqn\kone $$
It is also possible to prove \resgh\ directly for the $A(n,m,k)$ case,
in which $G/H =
 SU(n+m)_k / (SU(n) \times SU(m) \times U(1))$. The formula turns out
to be correct for all $n,m,k$, even
though there are fixed points when $n,m,k$ have a common factor.

\section{Fixed points}
When the field identification has fixed points, the correct
rule for Ramond ground states is to count one such state for
each identification orbit [\SchF]. In the analysis of the previous
subsection we have thus overcounted the non-fixed points by a factor
$p$, the order of the identification, and we still satisfied the
sum rule. This can be explained as follows. Let us assume for
the moment
that $p$ is prime, so that there are only simple fixed points.
The resolution of fixed points requires a matrix $\hat S$, defined only
on the fixed points, which together with the matrix $T$ of the fixed
points defines a representation of the modular group [\ScY].
In the case of $A(n,m,k)$ this
representation turns out to be precisely equal to the one of another
such theory, namely $A(n/p, m/p, k/p)$ [\SchF].
The charges of the "fixed point
conformal field theory" turn out to be ${1\over p}$ times the charges
of the corresponding fixed points of $A(n,m,k)$, and its central charge
is $\chat/ p^2$. Since the sum rule is
satisfied for the fixed point CFT, it follows that the fixed points
of $A(n,m,k)$ have an average charge-squared of precisely
$\chat/12$, and therefore counting them with any multiplicity, even
the wrong one, will not change the result.
If the fixed point CFT itself has field identification fixed points
one can just iterate this argument, so that we may conclude that
the average of the squares of the charges should be  equal to
${\chat \over{12}}$ for all sets of fixed points of the same order.

This ``fixed point independence"
is empirically also true for
KS-models of type $B$, $C$ and $D$, for which the fixed point
resolution matrix $\hat S$ cannot be identified with a known
$N=2$ theory. This can be understood in the following way. Before
one resolves the fixed point and normalizes the partition function,
one has a perfectly well-defined elliptic genus for the coset branching
functions. Since the partition function contains the identity more than
once the partition function and the elliptic genus do not correspond
to a sensible $N=2$ theory, but all the conditions for the sum rules
(essentially holomorphicity and  modular invariance) are satisfied.
This explains why even without dealing correctly with fixed points
one gets the correct average of $Q^2$. Furthermore the sum rule is
of course also satisfied if one deals correctly with the fixed points.
Hence the average of the squares of the
fixed point charges must separately be equal to ${\chat\over12}$.
Unlike the previous argument this one does not immediately generalize
to multiple fixed points, but in models of type $B$, $C$ and $D$
these do not occur anyway.

\section{Level-1 null vectors}
As a final application of the sum rules we show how they can be
used to get information about the null-vector structure of
$N=2$ $W$-algebras, information that would be rather difficult to
get directly from the algebra due to its non-linearity. The $N=2$
structure implies that extra generators must come in quartets
consisting of two bosonic generators of charge zero and two
fermionic ones of charge $\pm1$. Hence a Ramond ground state
with charge $q$ and $(-1)^F=1$ can have, in addition to its
$N=2$ excitations, four additional excited states generated by
the $-1$ modes of each such quartet. Supersymmetry requires these
states to come in pairs with opposite sign of $(-1)^F$, and charges
$(q, q+1)$ or $(q, q-1)$. These states appear unless they are null
states of the extended algebra. The presence of these states is
controlled by the first level sum rules, as explained in section 3.
If $|q| < {1\over 2} $ (which is always true if $\hat c < 1$)
each pair contributes with a negative sign, so that
no cancellations are possible. In this case the sum rules are
clearly more powerful than for $\chat > 1$, which is unfortunate
since this is precisely where extensions of the algebra become
important. Although the sum rules impose non-trivial restrictions
also for $\chat > 1$, we will restrict ourselves here to the
simpler case
$\chat \leq 1$.

\def\union{\cup}

Consider first $A(2,2,1)$. This theory has $\chat = {4\over 5}$ and
hence belongs to the minimal series. It is in fact a D-type
invariant of $A(1,1,8)$, whose chiral   algebra is extended
by the  simple current of $SU(2)_8$ and its $N=2$ partners.
Hence the $A(2,2,1)$ charges are a subset of
those of $A(1,1,8)$, and are $\pm {2\over5}$, $\pm{1\over5}$ each
occurring once
and $0$
occurring twice. At the first excited level the following $N=2$
doublets appear:
$$ \eqalign{
&a\times(-{2\over5},-{7\over5}) \union
   (1+b)\times(-{2\over5},{ 3\over5}) \union
   (1+c)\times(-{1\over5},{4\over5})\cr \union
  &(1+d)\times(-{1\over5},-{6\over5}) \union
   (1+e)\times(0,1)                  \union
   (1+f)\times(0,-1)                        + \hbox{c.c} \ .\cr} $$
The first member of each pair contributes with $(-1)^F =1$ and the
second with the opposite sign. The parameters $a\ldots f$ indicate the
unknown excitations due to the extra current quartet. Since there is
only one such quartet, these coefficients can only be 0 or 1, but
one does not actually need this limitation to solve the equations.
The
excitations of the $N=2$ algebra itself are known and are as
indicated.

The quadratic sum rule at the first level is
$\Tr[(-1)^F J_0^2] = -2 \chat \Tr[(-1)^F] = -{48\over 5}$. This yields
the following equation for the unknown coefficients
$$ -42-2(9 a -b -3c-7d-5e-5f) = -48 \ . $$
The only solution is $c=1$, and all other parameters equal to zero.
The quartic sum rule may be used as a check. If one does not restrict
the range of the parameters to 0 and 1, the quartic sum rule fixes
any remaining ambiguity.
It is interesting
that most of the expected excitations are null states.

A second example we have studied is $A(3,2,1)$, which has
$\chat=1$ and 10 chiral primary fields with charges
$\pm{1\over 2},\pm{1\over 3},2 \times (\pm {1\over 6})$  and $2\times 0$.
The
excited state can be described by 8 integer parameters. In this case
the quadratic equation has many degeneracies, but the quartic one
determines all parameters except one that cannot be determined by
any sum rule, namely
the multiplicity of a charge $(-{1\over 2},{1\over 2})$
pair.
Apart from the known $N=2$
excitations, we find the following pairs at the first level:
$(-{1\over 3},{2\over 3}),(-{1\over 6},{5\over 6}), (0,1) $
plus an unknown number of $(-{1\over 2},{1\over 2})$ pairs, plus complex
conjugates.

Although this information about first level excitations may not be
extremely important in itself, it does give interesting information
about the null state structure of $N=2$ W-algebras, which apparently
is rather subtle. If on the other hand one could understand the
null state structure directly from the algebra, one could use the
sum rules as a classification tool for the chiral rings that may
belong to such an algebra. In that case one would start with a set
of chiral primaries whose charges are the free parameters, and use
the sum rules to determine the central charge and the
allowed charges. For minimal models this would certainly give us
the first members of the series, if we had no other way of getting them.
Unfortunately our
present knowledge about W-algebras is too limited to contemplate
such a classification programme seriously.

\ack
A.N.S would like to thank the theory group
at the School of Physics and Astronomy, Tel-Aviv University, for
its hospitality.
\refout
\end
\bye
\bye